\documentclass[reprint,showpacs,twocolumn,superscriptaddress,aps,prb]{revtex4-2}
\usepackage{bbm}
\usepackage{bm}
\usepackage[most]{tcolorbox}
\usepackage{xcolor}
\usepackage{amssymb}
\usepackage{amsbsy}
\usepackage{mathtools}
\usepackage{amsmath}
\usepackage{stmaryrd}
\usepackage[dvipsnames]{xcolor}
\usepackage{graphicx}
\usepackage{soul}
\usepackage{epsfig}
\usepackage{placeins}
\usepackage{bbold}
\usepackage{braket}
\usepackage{blindtext}
\usepackage[colorlinks,linkcolor=blue,citecolor=blue,urlcolor=blue]{hyperref}

\usepackage{scalerel}
\usepackage{tikz}
\usetikzlibrary{calc}
\usetikzlibrary{patterns}
\usetikzlibrary{svg.path}
\definecolor{orcidlogocol}{HTML}{A6CE39}
\tikzset{
  orcidlogo/.pic={
    \fill[orcidlogocol] svg{M256,128c0,70.7-57.3,128-128,128C57.3,256,0,198.7,0,128C0,57.3,57.3,0,128,0C198.7,0,256,57.3,256,128z};
    \fill[white] svg{M86.3,186.2H70.9V79.1h15.4v48.4V186.2z}
                 svg{M108.9,79.1h41.6c39.6,0,57,28.3,57,53.6c0,27.5-21.5,53.6-56.8,53.6h-41.8V79.1z M124.3,172.4h24.5c34.9,0,42.9-26.5,42.9-39.7c0-21.5-13.7-39.7-43.7-39.7h-23.7V172.4z}
                 svg{M88.7,56.8c0,5.5-4.5,10.1-10.1,10.1c-5.6,0-10.1-4.6-10.1-10.1c0-5.6,4.5-10.1,10.1-10.1C84.2,46.7,88.7,51.3,88.7,56.8z};
  }
}

\newcommand\orcid[1]{\!%
  \href{https://orcid.org/#1}{%
    \mbox{%
      \scaleto{%
        \begin{tikzpicture}[yscale=-1,transform shape]
          \pic{orcidlogo};
        \end{tikzpicture}
      }{8pt}%
    }%
  }%
}

\newcommand{\mer}[1]{{\color{black} #1}}

\begin{document}

\title{
 Resonant level model from a Krylov perspective: \\  Lanczos coefficients in a quadratic model
}
\author{Merlin F\"{u}llgraf~\orcid{0009-0000-0409-5172}
}
\email{merlin.fuellgraf@uos.de}
\affiliation{University of Osnabr\"{u}ck, Institute for Physics, Barbarastr. 7, D-49076 Osnabr\"{u}ck, Germany}

 \author{Jiaozi Wang
\orcid{0000-0001-6308-1950}
 }
 \affiliation{University of Osnabr\"{u}ck, Institute for Physics, Barbarastr. 7, D-49076 Osnabr\"{u}ck, Germany}
\author{Jochen Gemmer
\orcid{0000-0002-4264-8548}
}
\affiliation{University of Osnabr\"{u}ck, Institute for Physics, Barbarastr. 7, D-49076 Osnabr\"{u}ck, Germany}%
\author{Stefan Kehrein
 \orcid{0009-0005-1119-4124}
 }
\affiliation{Georg-August-Universit\"{a}t G\"{o}ttingen, Institute for Theoretical Physics, Friedrich-Hund-Platz 1, D-37077 G\"{o}ttingen, Germany}
\date{\today}
\begin{abstract}

We study the Lanczos coefficients in a quadratic model given by an impurity interacting with a multi-mode field of fermions, also known as \textcolor{black}{resonant level} model.  We analytically derive closed expressions for the Lanczos coefficients of Majorana fermion operators of the impurity for different structures of the coupling to the hybridization band at zero temperature. While the model remains quadratic, we find that the growth of the Lanczos coefficients structurally depends strongly on the chosen coupling. Concretely, we find (i) approximately constant, (ii) exactly constant, (iii) square root-like as well as (iv) linear growth in the same model. 
We further argue that in fact through suitably chosen couplings, essentially arbitrary Lanczos coefficients can be obtained in this model.
These altogether evince the inadequacy of the Lanczos coefficients as a reliable criterion for classifying the integrability or chaoticity of the systems.
Eventually, in the wide-band limit, we find exponential decay of autocorrelation functions in all the settings (i)-(iv), which demonstrates the different structures of the Lanczos coefficients not being indicative of different physical behavior.

\end{abstract}
\maketitle

\section{Introduction}
Over several decades the recursion method \cite{viswanath94}, based on the Lanczos algorithm \cite{Lanczos1950zz}, has proven to be a highly versatile tool to tackle questions in quantum many-body physics \cite{viswanath90-rm,recursionmethod-applications}. Accelerated by the advent of the operator growth hypothesis (OGH)\ \cite{parker19}, the recursion method in operator space gained significant attention. Efforts along these lines encompass a \mer{wide} range of topics, including the study of transport properties \cite{uskov24-oleg-rm-2d,wang24,capizzi-rm-finite,bjoern-rm-apl-transport,lsl4-4lb4-rm-apl-green-function,mns1-l19c-markus-rm-dc,linz25}, equilibration times \cite{bartsch24,wang2024-2}, to the computation of correlation functions \cite{teretenkov25-rm-pseudomode,loizeau25-rm-sym,fuellgraf2025lanczospascalapproachcorrelationfunctions,PhysRevResearch.7.023245-rm-finite-temperature-dynamics,PhysRevB.110.155135-cao-rm-apl,shirokov2025-oleg-rm,ermakov2025symbolicrecursionmethodstrongly} and their dynamical features \cite{PhysRevE.111.024140-merlin-rm-arrow-of-time,teretenkov25-rm-pseudomode,fuellgraf2025lanczospascalapproachcorrelationfunctions}, Krylov complexities \cite{NANDY20251-review,rabinovici2025krylovcomplexity-krylov}, and many others \cite{menzler24-rm-chaos-apl,bhattacharya2022operator-ogh-open, bjoern-rm-apl}.

As a cornerstone of the recently developed recursion-method-based approach, the operator growth hypothesis postulates that,
in chaotic quantum many-body systems,
Lanczos coefficients of typical local observables exhibit asymptotically linear growth\ \cite{parker19}. 
Numerical simulations reveal the conjectured linear growth in a variety of chaotic systems\ \cite{noh21-ogh-num,heveling22-2}, while sublinear or bounded growth are typically observed in non-chaotic models, see e.g.\ \cite{viswanath94,lee-square-root,parker19} for square-root growth in interacting-integrable systems or \cite{parker19,NANDY20251-review,viswanath94,heveling22-2} for bounded growth in free models.

However, the connection between the chaoticity of a given system and the asymptotic growth of the Lanczos coefficients in it is more subtle.
Apart from chaotic systems as addressed in \cite{parker19}, linear growth has also been reported in certain non-chaotic settings, e.g.,
specific classically integrable systems with saddle-dominated scrambling \cite{bn-scrambling}. Notably these findings do not challenge the statement of the OGH, the intended domain of which concerns generic interacting chaotic systems, in which it postulates an asymptotic linear growth of the Lanczos coefficients related to local observables that do not couple to conserved quantities.

Further scrutinizing the interpretative power of the structure of the Lanczos coefficients regarding the chaoticity or integrability of the respective system,  we analyze a quadratic, hence exactly solvable, model, where an impurity is coupled to a set of hybridization modes, known in the literature as the  resonant level model. 
More precisely, we consider Majorana fermion operators of the impurity, and investigate the influence of different coupling structures on the structure of the respective Lanczos coefficients at zero temperature. 
For several different type of coupling structures, we are able to analytically derive \textit{all} Lanczos coefficients.
Despite the exact solvability of the model, we find markedly different behaviors of the Lanczos coefficients even at zero temperature, including ((i) approximately) (ii) constant, (iii) square-root, and (iv) linear growth.
Our findings suggest that the mere structure of the Lanczos coefficients is not necessarily correlated with the chaoticity/integrability of the underlying system. Furthermore, it sheds light on the possibility of growing Lanczos coefficients in the absence of operator growth.

In the remainder of the paper, we first revisit the framework of the recursion method in Sec.~\ref{sec-framework}, and then introduce the model and the observables in Sec.~\ref{sec-model}.
Sections~\ref{sec-analytical} and~\ref{sec-dynamics} constitute the main parts of the paper, where we provide analytical insights into the Lanczos coefficients of Majorana fermion operators (Sec.~\ref{sec-analytical}) and discuss impact of  their structure on the dynamics of the autocorrelation functions (Sec.~\ref{sec-dynamics}). 
Eventually we conclude the discussion in Sec.\ \ref{sec-conclusion} and discuss the findings.

The roadmap of Sec. ~\ref{sec-analytical} and~\ref{sec-dynamics} is as follows:
\begin{itemize}
    \item Exploiting the structure of the model, we derive an integrodifferential equation governing the dynamics of the autocorrelation function and explicitly establish the relation between the kernel and the coupling profile.
    \item Under certain assumptions, we infer the corresponding Lanczos coefficients from the continued-fraction expression of the kernel and obtain their explicit forms for four illustrative choices of the coupling density. Further, we argue for the generality of the findings based on an existence argument invoking Bochner's theorem.
    \item We then investigate how the structure of the Lanczos coefficients affects the dynamics by varying the width of the coupling profile/density, with particular emphasis on the wide-band limit.
\end{itemize}

\section{Recursion Method Framework\label{sec-framework}}
Let us consider the operator space, with its vectors written as $\vert\mathcal{O})$. The space is equipped with an inner product denoted by $( - \vert -)$, which defines a norm via $\Vert {\cal O} \Vert = \sqrt{({\cal O}|{\cal O})}$.  
At inverse temperature $\beta$, a pertinent scalar product is defined as
\begin{equation}\label{def-inner-beta}
    ({\cal A}|{\cal B}):=\frac{1}{2}\text{tr}(\rho(\beta)({\cal A}^{\dagger}{\cal B}+{\cal B}A^{\dagger}))
\end{equation}
with $\rho(\beta)=e^{-\beta{\cal H}}/\text{tr}(e^{-\beta{\cal H}})$, with $\mathcal{H}$ as the system's Hamiltonian operator\ \cite{viswanath94,scalar-product}.
The time evolution is generated by the systems' Liouvillian superoperator $\mathcal{L}=\left[\mathcal{H},-\right]$ via
\begin{equation}
    |{\cal O}(t)) = \exp{(i{\cal L}t)}|{\cal O}).
\end{equation}

The Lanczos algorithm now gives rise to the so-called Krylov basis $\{\vert\mathcal{O}_n)\}_n$ which tridiagonalizes the Liouvillian ${\cal L}$.
Starting with a normalized \textit{seed} state $\vert\mathcal{O})=\vert\mathcal{O}_0)$,
we set $b_1 = \Vert {\cal L}\vert{\cal O}_0)
\Vert$ as well as $|{\cal O}_1) = {\cal L} |{\cal O}_0)/b_1$. Then, we iteratively compute 
\begin{align}\label{eq-Lanczos}
        \begin{split}
            |\widetilde{\cal O}_n) &= {\cal L} |{\cal O}_{n-1}) - b_{n-1} |{\cal O}_{n-2}) \, , \\
b_n &= \Vert \widetilde{\cal O}_n \Vert \, ,  \\
 |{\cal O}_n) &= |\widetilde{\cal O}_n)/b_n \, ,
        \end{split}
\end{align}
where the $b_n$ denote the respective \textit{Lanczos coefficients}.
Therefore, in the Krylov basis $|{\cal O}_n)$, ${\cal L}$ becomes
\begin{equation}
{\cal L}_{mn} = ({\cal O}_{m}| {\cal L} |{\cal O}_{n}) = \left(\begin{array}{cccc}
0 & b_{1} & 0 & \cdots \\
b_{1} & 0 & b_{2} \\
0 & b_{2} & 0 & \ddots \\
\vdots & & \ddots & \ddots
\end{array} \right)_{mn}.
\end{equation}
The autocorrelation function can  then be defined as
\begin{equation}\label{eq-lanczos-autocorrelation}
{\cal C}(t)=({\cal O}_{0}|\mathrm{e}^{i{\cal L}t}|{\cal O}_{0})=(\mathcal{O}_0\vert\mathcal{O}_0(t)) ,
\end{equation}
with its Laplace transform connecting to the Lanczos coefficients via a continued fraction representation
\begin{align}
{\cal C}(s) := \mathcal{LT}\left[{\cal C}(t)\right](s)
=
\frac{1}{\displaystyle
s+\frac{b_{1}^{\,2}}{\,s 
+\frac{b_{2}^{\,2}}{\,s
+\frac{b_{3}^{\,2}}{\,s+
\dots}}}} ,
\label{eq-bn-cont-frac}
\end{align}
see e.g.\ \cite{viswanath94}.
In the remainder of this paper, we focus on zero temperature $\beta \rightarrow \infty$, in which case the inner product in Eq.\ \eqref{def-inner-beta} becomes 
\begin{align}(\mathcal{A}\vert\mathcal{B})=\frac{1}{\text{2dim(GS)}}\sum_{\Omega\in\text{GS}}\langle\Omega\vert\mathcal{A}^\dagger\mathcal{B}+\mathcal{B}\mathcal{A}^\dagger\vert\Omega\rangle,
\end{align}
where $\text{GS}$ represents the ground state manifold.

\section{Model and observable.\label{sec-model} }
We consider the single impurity Anderson model with vanishing interactions on the impurity, or resonant level model, described by the Hamiltonian
\begin{align}
    \mathcal{H}=\sum_k \epsilon_k c^{\dagger}_{k} c^{}_k+\epsilon_d d^{\dagger}d+\sum_k V_k\left(c_k^\dagger d+d^{\dagger}c_k^{}\right),\label{eq-resonant-level}
\end{align}
where the $d^\dagger,d$ and $c_k^\dagger,c_k^{}$ are the creation and annihilation operators in the impurity and the band fermions respectively. The parameters $\epsilon_d, \epsilon_k$ indicate the corresponding energies. The parameters $V_k$ encode the coupling between the impurity and the hybridisation modes. Note that the model is quadratic and thus exactly solvable.

As observables of interest, we consider two single-particle observables, i.e., Majorana fermion operators of the impurity ${\cal O} = \gamma_{1},\ \gamma_2$, where $\gamma_{1}=d^{\dagger}+d$ and $\gamma_{2}=i(d^{\dagger}-d)$. The Lanczos coefficients of ${\cal O}$ can in principle be calculated using Eq.~\eqref{eq-Lanczos}.
However, in the model we consider here, it is more convenient to directly study the autocorrelation function
\begin{equation}
{\cal C}(t)=(\mathcal{O}\vert\mathcal{O}(t))=\frac{1}{2}\langle0|\left({\cal O}(t){\cal O}+{\cal O}{\cal O}(t)\right)|0\rangle,
\end{equation}
with $\vert0\rangle$ as the system's vacuum, focusing
 on zero temperature.
With straightforward derivations,
one gets 
\begin{equation}
    {\cal C}(t)=\begin{cases}
\Re[a_{d}(t)] & {\cal O}=\gamma_{1}\\
-\Re[a_{d}(t)] & {\cal O}=\gamma_{2}
\end{cases}.
\end{equation}
Here 
\begin{equation}
 a_d(t)=\langle0|de^{-i{\cal H}t}d^{\dagger}|0\rangle\ ,
\end{equation}
which is the excitation amplitude in the impurity, starting from the initial state $|\psi(0)\rangle = d^\dagger |0\rangle$.
In the following, we focus on the Majorana fermion operator ${\cal O} = \gamma_1$, and the results can be straightforwardly generalized to ${\cal O} = \gamma_2$.
Note that, since the model considered here is quadratic, single-particle operators remain confined to the single-particle sector under time evolution, implying the absence of operator growth (see the appendix \ref{app-no-growth} for more detailed discussion).

\section{Analytical insights to the Lanczos coefficients\label{sec-analytical} }
Employing standard techniques as in the analysis of the structurally related Weisskopf-Wigner model in quantum optics, see e.g.\ \cite{Scully_Zubairy_1997} one may write for the state vector $\vert\psi\rangle$ of the system 
\begin{align}
    \vert\psi(t)\rangle=a_{d}(t)d^\dagger\vert0\rangle+\sum_k a_k(t)c_k^\dagger\vert 0\rangle,
\end{align}
where the initial conditions
\begin{align}
    a_d(0)=1,\quad a_k(0)=0,\label{eq-init-condition}
\end{align}
describe the scenario of an excitation being initialized in the impurity. Upon shifting the overall energy scale in the model such that $\epsilon_d\rightarrow\epsilon^\prime_d=0,$
it follows for the amplitude in the impurity that
\begin{align}
    \dot{a}_d(t)&=-\sum_k\vert V_k\vert^2\int_{0}^t dt^\prime e^{-i\epsilon_k(t-t^\prime)}a_d(t^\prime)\\
    &=-\int_0^t \mathcal{K}(t-t^\prime)a_d(t^\prime),\label{eq-kernel-equation}
\end{align}
where 
\begin{align}
    \mathcal{K}(t)=\sum_k \vert V_k\vert^2 e^{-i\epsilon_kt}.
\end{align}

\begin{table}[t]
\begingroup
\renewcommand{\arraystretch}{2.5} 
    \begin{tabular}{ |c |c | c| c|}
    \hline
    & $J(\omega)$ &  $\mathcal{K}(t)/\mu$& $\beta_n$\\
     \hline
     \hline
i&   $\frac{\mu}{\alpha}\sqrt{\frac{\pi}{2}}\Theta\left(\alpha^2-\omega^2\right)$&$\frac{\sin(\alpha t)}{\alpha t}$&$\frac{\alpha n}{\sqrt{(2n-1)(2n+1)}}$\\
    \hline
   ii& $\mu\frac{\sqrt{4\alpha^2-\omega^2}}{\sqrt{2\pi}\alpha^2}\Theta\left(4\alpha^2-\omega^2\right)$&$\frac{J_1(2\alpha t)}{\alpha t}$&$\alpha$\\
     \hline
 iii&  $\frac{\mu}{\alpha}\exp\left(-\frac{\omega^2}{2\alpha^2}\right)$&$\exp\left(-\frac{(\alpha t)^2}{2}\right)$&$\alpha\sqrt{n}$\\
     \hline
 iv&  $\frac{\mu}{\alpha}\sqrt{\frac{\pi}{2}}\text{sech}\left(\frac{\pi}{2\alpha}\omega\right)$&$\text{sech}\left(\alpha t\right)$&$\alpha n$\\
     \hline
     \end{tabular}
     \endgroup
 \caption{Correspondence table between the coupling density $J(\omega)$, the (rescaled) kernel $\mathcal{K}(t)/\mu$ (given as the Fourier transform of the former) and the Lanczos coefficients $\beta_n$ pertaining to $\mathcal{K}(t)/\mu$ for four different cases. These cases indicate $J(\omega)$ being given by a coupling to (i) hard-edge box, (ii) semi-circle, (iii) Gaussian and (iv) hyperbolic secant profile.}
 \label{tab:Summary}
\end{table}

We assume that the modes are dense and that the density of states in the band is uniform, such that
\begin{align}
    \sum_k\rightarrow \int d\omega\ \rho(\omega), \quad\text{s.t.}\ \vert V_k\vert^2 \rho(\omega)\rightarrow\vert V(\omega)\vert^2 \rho(\omega),
\end{align}
Defining a \textit{coupling density} $J(\omega)=\sqrt{2\pi} \rho(\omega)\vert V(\omega)\vert^2$, we may rewrite the kernel as 
\begin{align}
    \mathcal{K}(t)=\int d\omega\frac{J(\omega)}{\sqrt{2\pi}}e^{-i\omega t}.\label{eq-fourier}
\end{align}
i.e.\ as the Fourier transform of the coupling density $J(\omega)$.
With the kernel at hand we may write for Eq.\ (\ref{eq-kernel-equation}) 
\begin{align}
    \mathcal{LT}\left[a_d(t)\right](s)=:a_d(s)=\frac{1}{s+\mathcal{K}(s)},\label{eq-laplace-ode}
\end{align}
with $\mathcal{K}(s)$ denoting the kernel's Laplace transform.

Comparing Eqs.\ (\ref{eq-bn-cont-frac}) and\ (\ref{eq-laplace-ode}) we deduce that analyzing the Laplace transform of the kernel comprises all information about the Lanczos coefficients of the function $a_d(t)$. Hence we follow this approach as opposed to iteratively computing the Lanczos coefficients via the algorithm (\ref{eq-Lanczos}).

\begin{figure*}[t]
  \includegraphics[width=1.\textwidth]{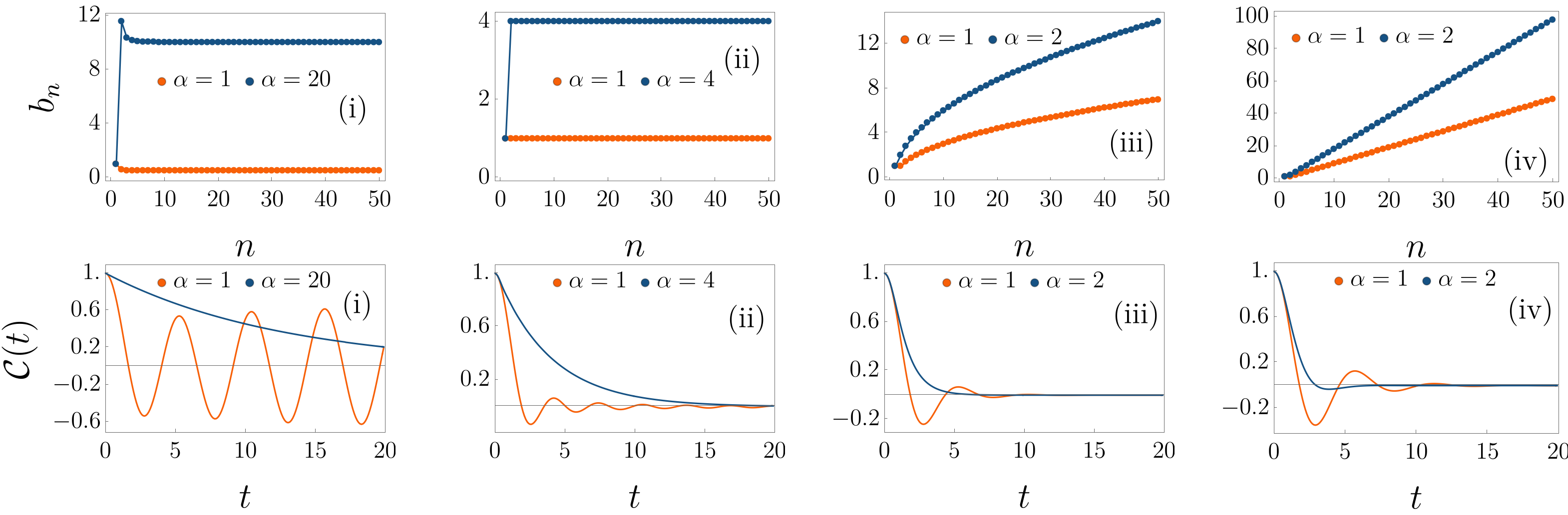}
  \caption{Exemplary Lanczos coefficients and autocorrelation functions for the couplings in Tab.\ \ref{tab:Summary} for different values of $\alpha$. For all cases we set $\mu=1$. \label{fig-sketch}}
\end{figure*}

We turn to four different coupling structures $J(\omega)$ specifically, for which we analytically determine \textit{all} Lanczos coefficients pertaining to the Majorana fermion operator $\mathcal{O}=\gamma_1 = d+d^\dagger$. Concretely, we study the cases where the $J(\omega)$ models a coupling of the impurity to (i) a hard-edge box, (ii) a semi-circle, (iii) a Gaussian and eventually (iv) a hyperbolic secant profile. The respective coupling densities are given in Table\ \ref{tab:Summary}. The latter give rise to the respective memory kernels via a Fourier transform, cf.\ Eq.\ (\ref{eq-fourier}). From Eq.\ (\ref{eq-laplace-ode}) it becomes evident that their Laplace transform, or rather the continued fraction representation thereof, will fully determine the dynamics of $\mathcal{C}(t)$. In fact, the (rescaled) kernels (see Tab.\ \ref{tab:Summary}) themselves allow for continued fraction representations of their Laplace transforms, see Tab.\ \ref{tab:Summary}. These connections have already been mathematically established in Ref.\ \cite{viswanath94,Joslin86}, albeit apart from the connection to a physical model. With (i) being the least discussed scenario in the literature, in App.\ \ref{app-derivation-box-bn} we provide a derivation for the Lanczos coefficients.

Revisiting Eq.\ (\ref{eq-laplace-ode}), the coefficients $\beta_n^{(a)}$, with $a$ indicating the specific coupling, allow us to straightforwardly infer the Lanczos coefficients $b_n^{(a)}$ pertaining to $\mathcal{C}(t)$ as
\begin{align}
    b^{(a)}_n=\begin{cases}
        \sqrt{\mu}\quad&\text{for}\ n=1,\\
        \beta^{(a)}_{n-1}\quad&\text{for}\ n\ge2.
    \end{cases}\label{eq-bn-beta-amplitude}
\end{align}

See Fig.\ \ref{fig-sketch} for an impression of the $b_n$ and corresponding dynamics for the cases considered in Tab.\ \ref{tab:Summary}.
Despite considering the four cases given in Tab.\ \ref{tab:Summary} explicitly, we note that such an analysis in fact may be readily generalized. To this end, consider some set of Lanczos coefficients $\{\beta_n^\prime\}_n$ together with its corresponding autocorrelation function $\mathcal{K}^\prime(t)$. Expressing the Laplace transform of the latter in its continued fraction representation and plugging it into Eq.\ (\ref{eq-laplace-ode}), we get for $\mathcal{C}(t)$ the Lanczos coefficients $b_1=1$ and $b_{n\ge2}=\beta^\prime_{n-1}$. That is, by choosing a particular memory kernel $\mathcal{K}^\prime$ we may bequest its Lanczos coefficients directly onto $\mathcal{C}(t)$.
Note that, as $\mathcal{K}^\prime (t)$ itself is an autocorrelation function, by Bochner's theorem there exists some $J^\prime(\omega)\ge0$, symmetric around $\omega=0$, such that an equation of the form of Eq.\ (\ref{eq-fourier}) holds and consequently allows for the interpretation a respective coupling density. 

It is worth mentioning that while this argument establishes the existence of a suitable coupling density $J^\prime(\omega)$, it may be that the latter is not easily experimentally realizable.
However, on the theoretical side, for \textit{any} ``desired" form for the Lanczos coefficients amounting to an autocorrelation function $\mathcal{K}^\prime(t)$, there \textit{exists} a coupling density giving rise to them, all leaving the free structure of the system unaltered. Therefore it is evident, that the asymptotic behavior of the Lanczos coefficients cannot be any indicator for the system at hand being either free, interacting-integrable or chaotic. 

Further, with $\mathcal{H}$ quadratic, a single-particle operator such as $\mathcal{O}$ will remain a single-particle operator under time-evolution. In this sense, in the setting considered here, there is no operator growth, despite Lanczos coefficients not being bounded. For more details see App.\ \ref{app-no-growth}.

\section{Impact on the dynamics\label{sec-dynamics}}
\begin{figure}
    \centering
    \includegraphics[width=0.8\linewidth]{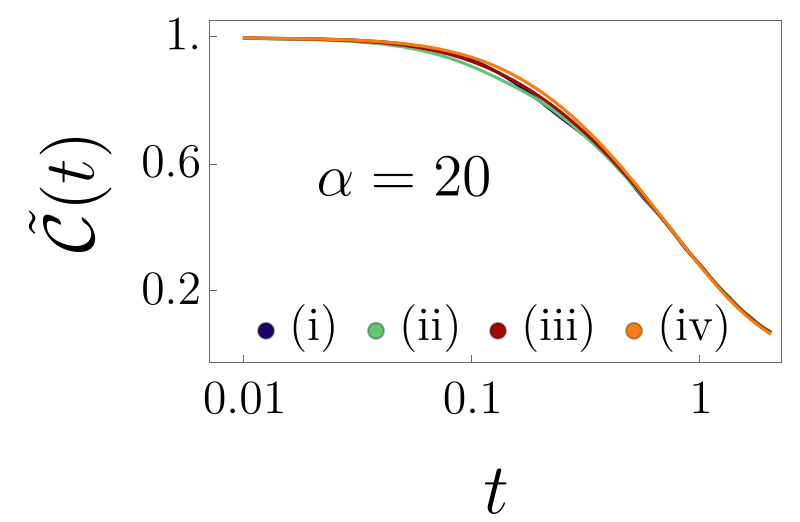}
    \caption{Dynamics of the autocorrelation function $\tilde{\mathcal{C}}(t)$ from the rescaled couplings with $\alpha=20$ depicted for short times.}
    \label{fig:zoomed}
\end{figure}

Finally, we investigate the impact of the structure of the Lanczos coefficients imposed by the different coupling structures discussed for the cases (i)-(iv) onto the eventual dynamics of the autocorrelation function. For the sake of better clarity we rescale the coupling densities $J^{a}(\omega)$ in order to render them comparable with respect to width and strength. The former are essentially governed by the parameter $\alpha$ for
\begin{align}
  \begin{split}
        \Tilde{J}^{(i)}(\omega)&=J^{(i)}(\omega)\vert_{\mu=\alpha\sqrt{2/\pi}},\\
    \Tilde{J}^{(ii)}(\omega)&=J^{(ii)}(\omega)\vert_{\mu=\alpha\sqrt{\pi/2}},\\
    \Tilde{J}^{(iii)}(\omega)&=J^{(iii)}(\omega)\vert_{\mu=\alpha},\\
    \Tilde{J}^{(iv)}(\omega)&=J^{(iv)}(\omega)\vert_{\mu=\alpha\sqrt{2/\pi}}
  \end{split}\label{eq-rescaled-coupling}
\end{align}
Such rescaling will merely affect the prefactors of the respective Lanczos coefficients, but leaves the structure thereof unaltered.
\begin{figure}[h]
    \centering
    \includegraphics[width=1\linewidth]{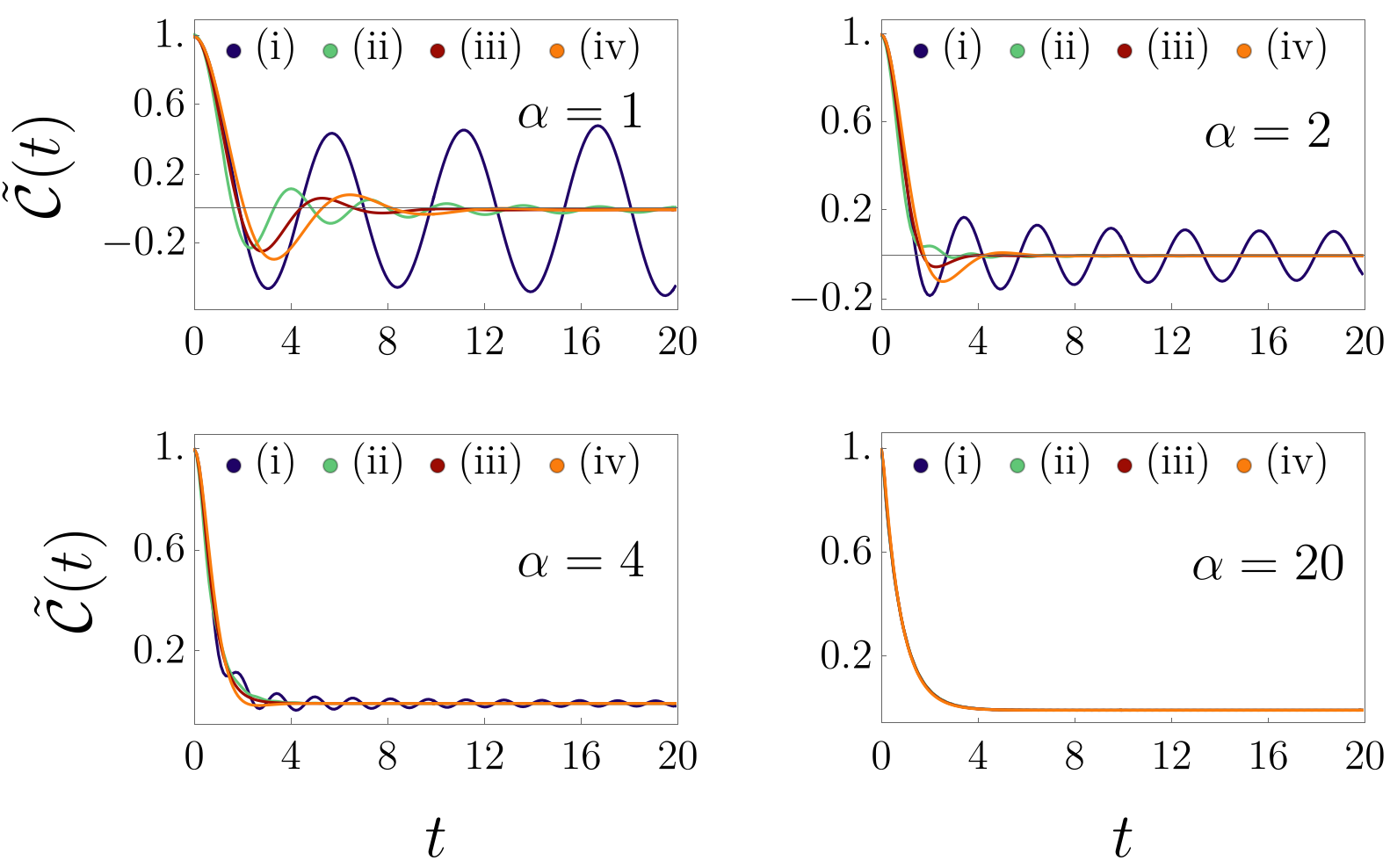}
    \caption{Dynamics from the rescaled coupling densities given in Eq.\ (\ref{eq-rescaled-coupling})  for different $\alpha$.}
    \label{fig:placeholder}
\end{figure}
In Fig.\ \ref{fig:placeholder} we depict the dynamics obtained for the autocorrelation function, denoted by $\tilde{\mathcal{C}}(t)$, for the different cases (i)-(iv) for different values of $\alpha$ by numerically evaluating the inverse Laplace transform given in Eq.\ (\ref{eq-laplace-ode}). 
Despite the qualitatively different structure of $b_n$, the distinctions in $\tilde{\cal C}(t)
$ among the different cases become less pronounced as $\alpha$ increases. For sufficiently large $\alpha$, such as $\alpha = 20$, the differences remain visible only at very short times (Fig.\ \ref{fig:zoomed}).

To better understand this behavior, let us consider the hierarchy of relevant timescales. The bandwidth $\alpha$ determines the bath correlation time, $\tau_B \sim 1/\alpha$, while the relaxation timescale of the impurity scales as $\tau_R \sim \frac{1}{J(0)} \sim \frac{\alpha}{\mu}$. 

Therefore, in the limit $\alpha \to \infty$, one has $\tau_R \gg \tau_B$. This corresponds to the Markovian limit, which in the case, leads to ${\cal C}(t) \propto e^{-\Gamma_R t}$ with $\Gamma_R \propto \frac{\mu}{\alpha}$ .

Here, with the rescaled coupling densities from Eq.\ (\ref{eq-rescaled-coupling}), we have $\widetilde{J}(0)=\text{const.}$ and consequently $\widetilde{C}(t)\propto e^{-\widetilde{\Gamma}_R t}$, with $\widetilde{\Gamma}_R=\text{const.}$
From that analysis we infer that in the wide-band limit ($\alpha\rightarrow\infty$) the very different behavior of the Lanczos coefficients is not indicative of different physical behavior. 
See App.\ \ref{app-krylov} for an investigation of the respective Krylov complexity in the cases (i)-(iv).

\section{Conclusion and discussion\label{sec-conclusion}}
In this paper we investigated the Lanczos coefficients of a Majorana operator in the resonant level model;
 a setup in which a non-interacting impurity is coupled to a band of fermionic hybridisation modes. For several kinds of coupling structures we analytically derived closed forms for the Lanczos coefficients of Majorana fermion operators of the impurity. These range from constrained, constant coefficients to square-root and linear growth. Given that the considered model is quadratic, these findings clearly demonstrate that the structure of the Lanczos coefficients does not allow to judge on the chaoticity or integrability of the examined system. We additionally argued that by an appropriate choice of the coupling, essentially arbitrary behaviors for the Lanczos coefficients may be obtained. Moreover, with the model being quadratic, these findings evince the possibility of growing Lanczos coefficients in the absence of operator growth.
Lastly, by analyzing the influence of the Lanczos coefficients on the occupation dynamics we demonstrate that in the wide-band limit of the coupling the resulting dynamics are similarly not indicative of different physical behavior, albeit stemming from structurally different Lanczos coefficients.
Further investigations regarding the connection interlinking chaoticity, Lanczos coefficients and dynamical features, e.g.\ the simplicity of dynamics, manifest in exponential decays of the autocorrelation remain interesting perspective for future research. Furthermore, although adding Hubbard-type interactions to the model in Eq.\ (\ref{eq-resonant-level}), which then describes the single impurity Anderson model, prevents the analytical treatment discussed in this paper, studying this more general model is an interesting path for future research.

\section{Acknowledgements}
M.F. and J.W. thank Björn Sbierski and Jannis Eckseler for fruitful discussions.
M.F., J.W., and J.G. acknowledge funding by the Deutsche Forschungsgemeinschaft (DFG), under Grant No. 531128043, as well as under Grant No. 397107022 within the DFG Research Unit FOR 2692, under Grant No. 355031190.
S.K. acknowledges support from the Deutsche Forschungsgemeinschaft (DFG) through FOR 5522, project T1 (Project-ID 499180199).

\section{Data availability} Research data associated with this article are openly available \cite{fullgraf_2026_19347500}.

\bibliography{main}
\clearpage
\appendix

\section{Lanczos coefficients for the box-coupling\label{app-derivation-box-bn}}

In order to derive the Lanczos coefficients for the box coupling given in case (i), see Tab.\ \ref{tab:Summary}, we make use of the continued fraction representation of the corresponding kernel. First, consider the Laplace transform
\begin{align}
 \mathcal{K}(s)=\mathcal{LT}\left[\frac{\sin\left(\alpha t\right)}{\alpha t}   \right](s)=\frac{1}{\alpha}\arctan\left(\frac{\alpha}{s}\right).
\end{align}
The arctan allows for a continued fraction representation given by \cite{wolfram}
\begin{align}
   \frac{1}{\alpha} \arctan\left(\frac{\alpha}{s}\right)=\frac{\left(1/s\right)}{1+\frac{\left(\alpha/s\right)^2}{3+\frac{4\left(\alpha/s\right)^2}{5+\frac{9\alpha^2}{7s+\dots}}}}=\frac{1}{s+\frac{\alpha^2}{3s+\frac{4\alpha^2}{5s+\frac{9\alpha^2}{7s+\dots}}}}.\label{eq-app-cont-frac}.
\end{align}
In order to obtain a form as in Eq.\ (\ref{eq-bn-cont-frac}) we perform an equivalence transformation to the factors in the partial denominators. The resulting partial nominators will in turn yield the Lanczos coefficients related to $\mathcal{K}(t)$ by comparison with Eq.\ (\ref{eq-bn-cont-frac}). The former, denoted by $\beta_n$, are then found to yield 
\begin{align}
    \beta_n=\frac{\alpha n}{\sqrt{(2n-1)(2n+1)}}.
\end{align}

\section{Growth of Lanczos coefficients without operator growth\label{app-no-growth}}

In this section, we provide a discussion of the physical picture for growing Lanczos coefficients in the absence of operator growth
taking the example of ${\cal O} = \gamma_1 = d^\dagger + d$.

Note that the operator ${\cal O}$ is a single-particle operator.
In a quadratic Hamiltonian, a single-particle operator remains a single-particle operator throughout the time evolution
\begin{equation}
{\cal O}(t)=d^{\dagger}(t)+d(t)=u(t)\,d+u^{*}(t)\,d^{\dagger}+\sum_{k}\Bigl[\beta_{k}(t)\,c_{k}+\beta_{k}^{*}(t)\,c_{k}^{\dagger}\Bigr].
\end{equation}
This means that in the case we consider here, there is no operator growth.
This is also evident from the Lanczos algorithm: all Krylov basis operators are single-particle operators
\begin{equation}
    |{\cal O}_{n})=u^{(n)}\,d+u^{(n)*}\,d^{\dagger}+\sum_{k}\Bigl[\beta_{k}^{(n)}\,c_{k}+\beta_{k}^{(n)*}\,c_{k}^{\dagger}\Bigr]\ .
\end{equation}

When applying the Liouvillian operator $\mathcal{L}$ on $|\mathcal{O}_n)$, only the coefficients $u^{(n)},\ \beta_k^{(n)}$ change, while the effective dimension of the operator space remains the same (albeit infinite).
Nevertheless, as exemplified at hand of the coupling densities in the main text, the Lanczos coefficients $b_n$ may in fact grow, e.g.\ linearly as in case $(iv)$.

As a possible explanation for the linearly growing $b_n$ in this case, let us consider the following scenario. 
We apply ${\cal L}$ to the Krylov basis $|{\cal O}_{n})$ , i.e.\  $|{\cal \widetilde{O}}_{n+1})={\cal L}|{\cal O}_{n})$
If the following holds for all $n > n_0$
\begin{equation}
({\cal O}_{n-1}|{\cal \widetilde{O}}_{n+1})=0,\ \ ({\cal \widetilde{O}}_{n+1}|{\cal \widetilde{O}}_{n+1})=\alpha > 1 ,
\end{equation}
turning to the Lanczos coefficients one obtains
\begin{equation}
    b_n \sim \alpha n + b\ \   \forall n > n_0 .
\end{equation}

\section{Krylov complexity\label{app-krylov}}
\begin{figure}[h]
    \centering
\includegraphics[width=1\linewidth]{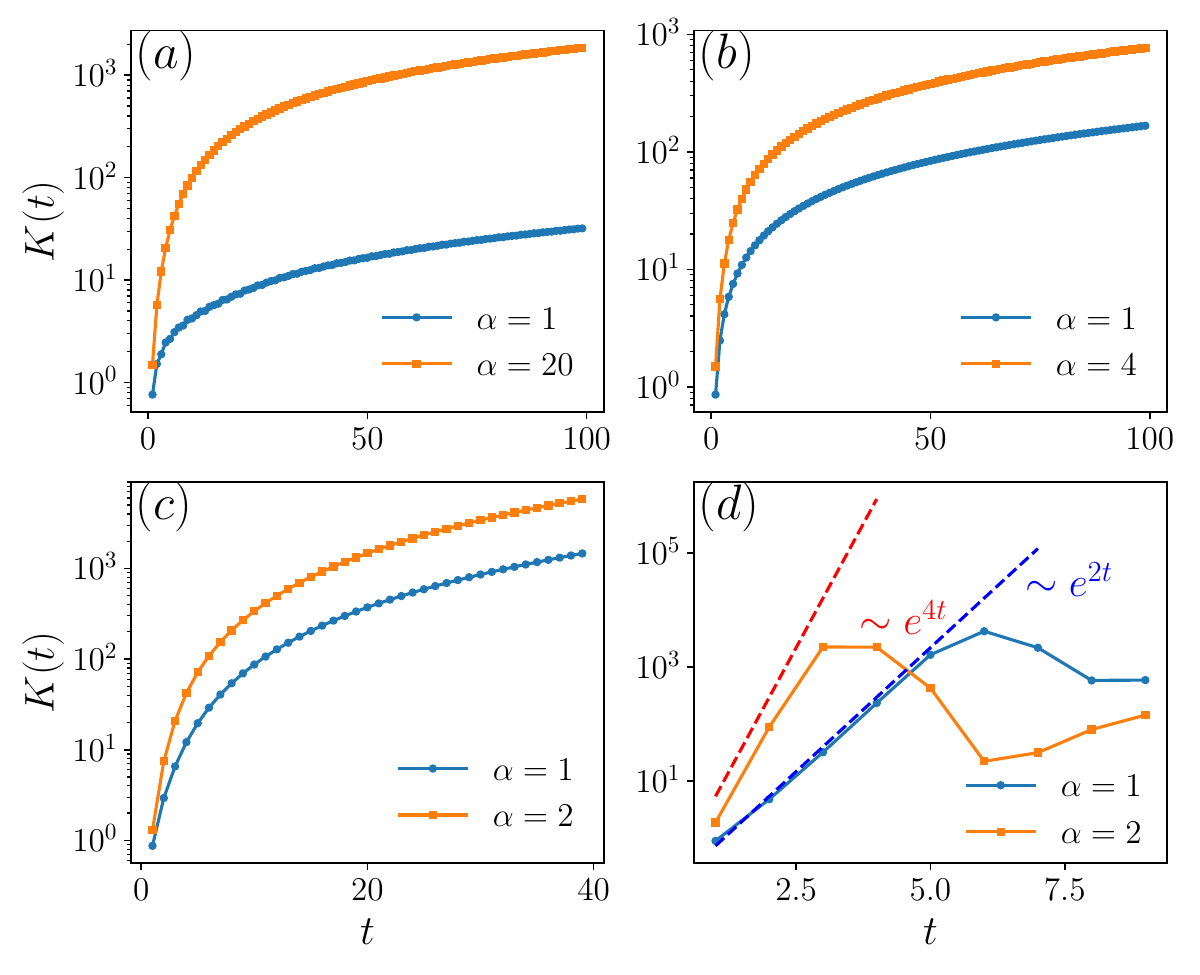}
  \caption{Krylov complexity $K(t)$ (in logarithmic scale) for the couplings in Tab.\ \ref{tab:Summary} for different values of $\alpha$. For all cases we set $\mu=1$. The simulation is done using a finite $d \times d$  Liouvillian operator, with $d = 10000$. The dashed lines in (d) indicate the scaling $\sim e^{2\alpha t}$.  \label{fig-KC}}
\end{figure}

In Fig.~\ref{fig-KC}, we present results for the Krylov complexity $K(t)$ for couplings in Tab.\ \ref{tab:Summary} for different values of $\alpha$.

\end{document}